\documentclass[a4paper,11pt]{article}
\usepackage[a4paper, total={17cm, 25cm}]{geometry}
\usepackage{fancyhdr}
\fancypagestyle{plain}{
\fancyhf{} 
\fancyhead[RH]{\thepage}
\fancyfoot[CF]{28th Texas Symposium on Relativistic Astrophysics \\ Geneva, Switzerland -- December 13-18, 2015}

}
\begin{document}

\title{\bf Unveiling the parent population of beamed \\
narrow-line Seyfert 1s}

\author{
Marco Berton, Universit\`a degli studi di Padova, Italy 
\\
Luigi Foschini, INAF - Osservatorio astronomico di Brera, Italy 
\\
Stefano Ciroi, Universit\`a degli studi di Padova, Italy 
\\
Alessandro Caccianiga, INAF - Osservatorio astronomico di Brera, Italy
\\
Bradley M. Peterson, Ohio State University, Ohio, USA
\\
Smita Mathur, Ohio State University, Ohio, USA
\\
Matthew L. Lister, Purdue University, Indiana, USA
\\
Jennifer L. Richards, Purdue University, Indiana, USA
\\
Enrico Congiu, Universit\`a degli studi di Padova, Italy 
\\
Valentina Cracco, Universit\`a degli studi di Padova, Italy 
\\
Francesco Di Mille, Las Campanas Observatory, Chile
\\
Michele Frezzato, Universit\`a degli studi di Padova, Italy 
\\ 
Giovanni La Mura, Universit\`a degli studi di Padova, Italy  
\\
Piero Rafanelli, Universit\`a degli studi di Padova, Italy 
}

\date{February 15, 2016}

\maketitle

\begin{abstract}
Narrow-line Seyfert 1 galaxies (NLS1s) are active galactic nuclei (AGN) recently identified as a new class of $\gamma$-ray sources. The high energy emission is explained by the presence of a relativistic jet observed at small angles, just like in the case of blazars. When the latter are observed at larger angles they appear as radio-galaxies, but an analogue parent population for beamed NLS1s has not yet been determined. In this work we analyze this problem by studying the physical properties of three different samples of parent sources candidates: steep-spectrum radio-loud NLS1s, radio-quiet NLS1s, and disk-hosted radio-galaxies, along with compact steep-spectrum sources. In our approach, we first derived black hole mass and Eddington ratio from the optical spectra, then we investigated the interaction between the jet and the narrow-line region from the [O III] $\lambda\lambda$4959,5007 lines. Finally, the radio luminosity function allowed us to compare their jet luminosity and hence determine the relations between the samples.
\end{abstract}


\section{Introduction}

One of the most intriguing classes in the vast zoology of active galactic nuclei (AGN) are narrow-line Seyfert 1 galaxies (NLS1s). First classified by Osterbrock \& Pogge (1985), they exhibit a low full width at half maximum (FWHM) of H$\beta$ (by definition below 2000 km s$^{-1}$), and a ratio [O III]/H$\beta <$ 3. Such characteristics, along with the high Eddington ratio \cite{Boroson92}, are often interpreted as consequences of the presence of a relatively low mass black hole (10$^6$-10$^8$ M$_\odot$) accreting fast. One explanation for the low black hole mass is the young age of these objects \cite{Mathur00}. Another possibility is that the broad-line region (BLR) of these sources is disk-like shaped \cite{Decarli08}. In this case, NLS1s would be AGN observed pole-on, in which the Doppler broadening due to the rotation is not visible and the permitted lines appear to be narrow. \par
NLS1s drew new attention with the discovery of $\gamma$-ray emission from a few of them with the Fermi Satellite \cite{Abdo09a}. This high-energy emission is typically detected in radio-loud sources with flat radio-spectra and high brightness temperatures \cite{Yuan08}, which reveal that these objects likely harbor a relativistic jet pointed toward the Earth. NLS1s then became the third class of AGN with a beamed jet, along with BL Lacs and flat spectrum radio quasars (FSRQs). According to the unified model, the misaligned counterparts, also known as the parent population, of BL Lacs and FSRQs are radio-galaxies \cite{Urry95}. For beamed NLS1s, whose properties were extensively investigated by \cite{Foschini15}, the nature of such population is still uncertain. In this work we will present a review of the researches published in recent papers by Berton et al. \cite{Berton15a,Berton15b,Berton16}, in which this parent population problem was investigated in several aspects. \par
A first hypothesis regarding parent sources was suggested in \cite{Foschini11}. A flat-spectrum radio-loud NLS1 (F-NLS1), observed under a different inclination, appears as a steep-spectrum radio-loud NLS1 (S-NLS1) with jets viewed at large angles. Such sources exist, but they are quite rare. In the complete sample of \cite{Yuan08}, out of 23 radio-loud NLS1s (RLNLS1s) 12 have a flat-spectrum and only 7 have a measured steep-spectrum. For geometrical reasons, the theoretical number of misaligned sources should be 2$\Gamma^2$ times the number of beamed objects, where $\Gamma$ is the bulk Lorentz factor. For a typical value of $\Gamma \sim 10$ \cite{Abdo09c}, the 12 beamed sources in \cite{Yuan08} should correspond to $\sim$2400 misaligned objects but, as mentioned before, only 7 can be found in the sample. It is clear from these numbers that something is missing. \par
It is possible that this lack of sources is due to a selection effect. Beamed sources are more luminous because of relativistic effects, and therefore visible at larger distances. But it is also possible that misaligned F-NLS1s appear as completely different sources, and hence are not easily recognizable as such \cite{Foschini11}. If, for instance, NLS1s are young, it is possible that they have not developed radio lobes yet, and that only a strongly collimated jet is present. If this is the case, at large inclination the diffuse emission coming from the jet would be weak and virtually undetectable, making the source appear as a radio-quiet NLS1 (RQNLS1). Conversely, if NLS1s are AGN observed pole-on with a flattened BLR, a different angle would increase the Doppler broadening, and the source would appear as a broad-line radio-galaxy (or narrow-line radio-galaxy when the line of sight intercepts the molecular torus). As NLS1s often hosted in disk galaxies \cite{Crenshaw03}, parent sources should share the same property. A suitable candidate would then be a disk-hosted radio-galaxy. \par
Finally, another option must be kept on the table. Many authors have suggested a link between NLS1s and compact steep-spectrum objects (CSS, e.g. \cite{Oshlack01, Komossa06, Yuan08, Caccianiga14, Gu15}). The latter are also often considered young radio-galaxies with the jets still confined within the host-galaxy and radio-lobes developed only on a small scale \cite{Fanti95}, and RLNLS1s are reminescent of them in several respects \cite{Gu15}. For this reason, it is possible that misaligned F-NLS1s can be identified with CSS, and in particular with those having a high-excitation spectrum (CSS/HERG). To understand which of these hypotheses is correct, we investigated the physical properties of four samples of candidates approaching this problem with different methods. \par

\section{Methods}
\subsection{Black hole mass}
To derive this physical property along with the Eddington ratio, we analyzed the optical spectra, obtained from the SDSS or with the Asiago Astrophysical Observatory. We derived the mass under the assumption of virialized system, deriving in type 1 AGN the second-order moment $\sigma$ of the H$\beta$ line, since $\sigma$ is relatively insensitive to inclination effects and BLR geometry \cite{Peterson11}, and therefore it is particularly helpful in our case. For type 2 AGN, we instead used as a proxy the width of the [O III] core component. The latter is connected to the stellar velocity dispersion \cite{Nelson96} and, as a consequence, to the black hole mass. In both cases, the use of lines instead of the continuum partially avoids contamination from the jet synchrotron emission, and should provide a less biased estimate of black hole mass. To derive finally the bolometric luminosity, and hence the Eddington ratio, we measured the line luminosities, focusing again on H$\beta$ and [O III], which are both directly connected with the AGN luminosity \cite{Greene10}. \\
\subsection{Lines decomposition}
We studied the [O III] lines profiles in 124 NLS1s, 68 radio-quiet and 56 radio-loud. The first sample was derived from SDSS DR7, the second was a collection of all known RLNLS1s with a suitable optical spectrum. A typical [O III] lines profile can be decomposed in two components. The first one represents the line core, which is usually associated with gas moving far away from the nucleus, whose motion is dominated only by the gravitational potential of the bulge. The second one is a blue wing, thought to be originated in outflows moving in the inner narrow-line region (NLR), possibly generated by the strong radiation pressure coming from the accretion disk, or via interaction with a relativistic jet \cite{Komossa08}. We decomposed each [O III] line using two Gaussians. The fitting was performed by means of an automatic procedure, and the errors estimation on each parameter via Monte-Carlo method. \\
\subsection{Luminosity function} 
The luminosity function (LF) is the volumetric density of sources as a function of their luminosity. Complete samples are required to compute it. We used this technique on the only two complete samples available in literature of F-NLS1s and CSS/HERGs, and on a control sample of FSRQs. The method we used was the same described in \cite{Urry84, Padovani92}. For each source, we evaluated the maximum volume at which it could be detected, V$_{max}$, testing then the evolution of the samples with a V/V$_{max}$ test. Then we derived the LF at 1.4 GHz by dividing the samples in luminosity bins. We later used the model of relativistic beaming described in \cite{Urry84}. This technique allowed us to add relativistic beaming to the parent candidate sample, comparing then the resulting model with the LF obtained for the beamed sample, and testing the unification of the two populations on a statistical basis. 

\section{Steep-spectrum radio-loud NLS1s}
In the first parent sample, S-NLS1s, we studied the black hole mass and Eddington ratio distributions. The values we found are in good agreement with several works already published in the literature. S-NLS1s have masses between 10$^6$-10$^{8.5}$ M$_\odot$. Such values are quite close to those found among F-NLS1s which, according to \cite{Foschini15}, are between 10$^{6.5}$-10$^{8.5}$ M$_\odot$. The Kolmogorov-Smirnov test (K-S) we performed seems to point out that these two mass distributions might be drawn from the same population. This result is particularly robust when statistically complete subsamples of sources are considered. For the Eddington ratio distibutions the result does not change, with typical values between 0.01 and 1 in both samples. This indicates that S-NLS1s are very likely misaligned F-NLS1s, in analogy with what happens for blazars and radio-galaxies. At large inclinations, the radio-lobes extended emission starts to dominate over that of the core, making the radio-spectrum steeper. These sources cannot anyway represent the entire parent population, since this sample by construction does not include obscured parent candidates (type 2 AGN). 

\section{Disk-hosted radio-galaxies}
Disk RGs can be type 1 parent sources only if the BLR has a flattened geometry. If this is not the case, a large FWHM in the permitted lines is indeed an indication of high black hole mass, not compatible with that of NLS1s. Obscured RLNLS1s instead are more likely to appear as narrow-line radio-galaxies hosted in a disk galaxy. In our work, we analyzed the black hole mass distribution of these sources, finding that disk-hosted RGs black holes span a large range of masses. This class of AGN appears to be similar to a bridge connecting the low-mass/high-Eddington sources, such as NLS1s, with high-mass/low-Eddington objects, such as elliptical radio-galaxies. Our finding is that those sources with low black hole mass and high Eddington ratio are suitable parent candidates, and that disk RGs with a pseudobulge can be even better candidates as F-NLS1s parent sources. \par
Our sample included a few type 1 AGN, which have FHWM larger than 2000 km s$^{-1}$, but black hole masses that overlap with the F-NLS1s distribution. Our method is relatively insensitive to BLR geometry, so the higher FWHM of these sources and the similar black hole mass might be a hint of a flattened component in the BLR. It is finally worth noting that two of these disk-hosted radio-galaxies are the only $\gamma$-ray emitters included in our samples. They are 3C 120, hosted in a disk galaxy \cite{Inskip10} and detected by \cite{Abdo10}, and IC 310, hosted in a lenticular galaxy \cite{Paturel03} and identified as a TeV emitter by \cite{Kadler12}. Both these sources are included in the 3rd Fermi catalog of $\gamma$-ray emitting AGN \cite{Ackermann15}. 

\section{Radio-quiet NLS1s}
The black hole mass analysis on RQNLS1s did not provide a clear result. Their distribution is slightly different with respect to that of F-NLS1s, with values between 10$^{5.5}$-10$^{8.5}$ M$_\odot$ and on average a lower mass. Conversely the Eddington ratio distribution is close to that of F-NLS1s. Therefore we investigated these sources also by means of their [O III] lines profile, trying to compare the dynamics of their NLR to that of RLNLS1s (both steep- and flat-spectrum). Our result was that the NLR of RLNLS1s typically shows stronger signs of perturbation, likely due to interaction with a relativistic jet. In particular, we found that the [O III] lines in RLNLS1s often show a large blueshift, corresponding to velocities higher than 150 km s$^{-1}$. This feature likely indicates a bulk outflowing motion of the NLR in these sources \cite{Komossa08}. This behavior is observed also among RQNLS1s, but the number of such blue outliers is much less frequent, as confirmed by the Anderson-Darling statistical test we performed. The interpretation we gave for these results is that in RLNLS1s a fully developed, perhaps intermittent, relativistic jet is moving through the medium and interacting with it. In RQNLS1s, however, only a jet-base \cite{Falcke99} or an aborted jet \cite{Ghisellini04} is present, originating weak non-thermal emission in the core \cite{Giroletti09} and a strong wind which, despite being able to induce turbulence in the NLR, has not the same efficiency as the relativistic jet. Therefore the vast majority of RQNLS1s are probably not part of the parent population, though few exceptions are present (e.g. Mrk 1239 \cite{Doi15}). The latter are anyway likely due to the difficult measurement of the radio-loudness for nearby sources. Radio-loudness in fact is strongly affected by the way in which the optical and radio fluxes are evaluated. In particular, some sources can move from the radio-quiet to the radio-loud domain when in the optical flux only the nuclear component is considered \cite{Ho01}.

\section{Compact steep-spectrum sources}
To understand whether CSS, particularly those classified as HERG, are part of the parent population, we investigated again their black hole mass and Eddington ratio distribution. In this case we found a nearly perfect agreement between this class of sources and F-NLS1s, both in mass and in accretion luminosity. The K-S test indeed suggested that the two distributions are very likely originated by the same population. To get further confirmation of such result, we investigated CSS by means of the radio LF at 1.4 GHz. After computing it, as mentioned before we added a model of relativistic beaming to the parent candidate LF, and compared the result with the observed LF of F-NLS1s. We found that, using reasonable beaming parameters, such as bulk Lorentz factor $\Gamma = 10$ and ratio between jet luminosity and unbeamed luminosity $f = 0.5$, CSS/HERGs represent a very good candidate as parent population, at least in a statistical sense. It is worth noting that this does not contradict the previous conclusions on the parent sources. Both S-NLS1s and disk RGs are indeed probably connected with CSS/HERGs. As pointed out by several authors (see for instance the recent works by \cite{Caccianiga14} and \cite{Gu15}) S-NLS1s often show a radio morphology which is consistent with that of CSS. Moreover other studies (e.g. \cite{Morganti11, Best12}) showed that CSS/HERGs might be hosted in disk galaxies, therefore CSS/HERGs might be part of the the disk RGs class. Finally, the LFs seem also to point out a possible evolutionary scenario in which F-NLS1s and CSS/HERGs are the young and still growing version of FSRQs and Fanaroff-Riley radio-galaxies with high-excitation spectrum, respectively \cite{Berton16}. 

\section{Conclusions}
All the results we have previously summarized seem to unveil a new unified model of young AGN. This investigation led us to understand that, when observed at larger angles, F-NLS1s can appear as CSS objects with high-excitation spectra. Both S-NLS1s and disk RGs can be, with some caveats, considered as sources related to CSS/HERGs. Therefore the following picture seems to come to light. A beamed NLS1, when observed at larger angles without obscuration, appears as a S-NLS1, and can be classified as a CSS according to its radio morphology and young age. When the line of sight intercepts the molecular torus, the source appears as a disk-hosted narrow-line radio-galaxy, again with a CSS morphology. The incidence of the BLR geometry on this model is still unclear. More detailed studies on larger samples are then required to better address this problem, which might also reveal in turn very interesting information on the evolution of AGN through cosmic time. 


\end{document}